\documentclass[%
 reprint,
 preprintnumbers,
 amssymb,
 aps,
]{revtex4-1}

\usepackage{amsmath}
\usepackage[dvipdfmx]{graphicx}% Include figure files
\usepackage{dcolumn}% Align table columns on decimal point
\usepackage{bm}% bold math
\usepackage{braket}

\usepackage{ulem}
\usepackage{color}

\newcommand{\Slash}[1]{{\ooalign{\hfil/\hfil\crcr$#1$}}}

\begin{document}
\preprint{}
%\linenumbers

\title{
Model identification in $\mu^-\to e^-$ conversion\\ with invisible boson emission using muonic atoms
}% Force line breaks with \\
%\thanks{}%

\author{Yuichi Uesaka$^{1,2}$}
\affiliation{
$^1$Faculty of Science and Engineering, Kyushu Sangyo University, 2-3-1 Matsukadai, Higashi-ku, Fukuoka 813-8503, Japan and\\
$^2$Physics Department, Saitama University, 255 Shimo-Okubo, Sakura-ku, Saitama, Saitama 338-8570, Japan
}
% \altaffiliation[Also at ]{Department of Physics, Osaka University.}%Lines break automatically or can be forced with \\
%\author{Second Author}%
% \email{Second.Author@institution.edu}
%\affiliation{%
% Authors' institution and/or address\\
% This line break forced with \textbackslash\textbackslash
%}%

%\collaboration{}%\noaffiliation

%\author{}
% \homepage{}
%\affiliation{
% Second institution and/or address\\
% This line break forced% with \\
%}%
%\affiliation{
% Third institution, the second for Charlie Author
%}%
%\author{Delta Author}
%\affiliation{%
% Authors' institution and/or address\\
% This line break forced with \textbackslash\textbackslash
%}%

%\collaboration{CLEO Collaboration}%\noaffiliation

\date{\today}% It is always \today, today,
             %  but any date may be explicitly specified

\begin{abstract}
In this article, we investigate the $\mu^-\to e^-X$ process in a muonic atom, where $X$ is a light neutral boson.
By calculating the spectrum of the emitted electron for several cases, we discuss the model-discriminating power of the process.
We report the strong model dependence of the spectrum near a high-energy end point.
Our findings show that future experiments using muonic atoms are helpful to identify the properties of exotic bosons.
\end{abstract}

% PACS, the Physics and Astronomy
                             % Classification Scheme.
%\keywords{Suggested keywords}%Use showkeys class option if keyword
                              %display desired
\maketitle

\onecolumngrid

%\tableofcontents

\section{Introduction \label{sec:Intro}}

Though the standard model (SM) of particle physics is consistent with almost all experimental data, it still leaves many unanswered questions: the existence of dark matter, the origin of the neutrino masses, and so on.
To build physics beyond the SM, physicists have searched for direct or indirect clues for many years.
Since we have many candidates for the SM extension, we need to try various complementary methods to probe the effects of new physics.
Interestingly, several candidates predict light particles that interact feebly with the SM particles.
For the feebly interacting light particles, it is preferable to take a different approach from heavy particle searches.

If there is such a neutral boson $X$ with a mass smaller than a muon mass $m_\mu=105.658$ MeV, the boson $X$ induces an exotic muon decay $\mu\to eX$.
In fact, some promising phenomenological models include a new particle whose mass is of MeV or less and which induces the lepton flavor violation: e.g. light scalars such as majorons, familons, and axionlike particles \cite{Pilaftsis1994,Hirsch2009,Garcia-Celt2017,Wilczek1982,Celis2014,Celis2015,Ema2017,Jaeckel2014}, or light extra gauge bosons \cite{Dobrescu2005,Heeck2015,Heeck2016,Farzan2016}.
To investigate them generally, the authors of Ref.~\cite{Heeck2017} carried out a comprehensive study of $\ell\to\ell'X$ processes where the emitted $X$ decays into lighter SM particles like an electron-positron pair or a photon pair.

Let us consider cases where the $X$ has a sufficiently long lifetime or decays into invisible particles.
The general searches for the two-body muon decay $\mu^+\to e^+X$ have been performed in some experiments.
Even if we do not care about the decay property of the $X$, we can search for its trace by careful measurement of a positron energy spectrum in the muon decay.
Let $m_X$ be the mass of $X$, and you find the spectrum enhanced at $E_e\simeq\left(m_\mu^2-m_X^2\right)/\left(2m_\mu\right)$.
An inevitable background on this kind of search is positrons emitted from the ordinary muon decay, $\mu^+\to e^+\nu_e\overline{\nu}_\mu$, which is especially serious for a small $m_X$.
To suppress this background, the authors of Ref.~\cite{Jodidio1986} accumulated $1.8\times 10^7$ polarized positive muons and counted emitted positrons in the opposite direction to the polarization of muons.
As a result, they concluded that the constraint for the branching ratio was $Br(\mu^+\to e^+X)<2.6\times 10^{-6}$, assuming that the momentum distribution of signal positrons is spherically symmetric and the $X$ is massless.
Under this assumption, this constraint is still more stringent than those of any other experiments.
In 2015, the TWIST experiment~\cite{Bayes2015} reported the latest search for $\mu^+\to e^+X$.
They analyzed $5.8\times 10^8$ muons and obtained the branching ratio limits of $\mathcal{O}\left(10^{-5}\right)$ for various decay asymmetries and masses of 13 MeV$<m_X<$80 MeV.
In the near future, Mu3e Collaboration is going to investigate $\mu^+\to e^+X$ with sensitivity of $Br\sim \mathcal{O}\left(10^{-8}\right)$.
According to \cite{Perrevoort2018a,Schoening2017,Perrevoort2018b}, the explorable mass region of the search is 25 MeV$< m_X<$95 MeV.
This lower restriction comes from the difficulty of calibration due to the steep edge of the background spectrum, and the significant update of the constraints for $m_X\lesssim 25$ MeV would be challenging.

A different method to investigate the $\mu\to eX$ process is to use muonic atoms instead of free muons, which was proposed in Ref.~\cite{Tormo2011}.
According to the literature~\cite{Tormo2011}, coming experiments using muonic atoms, such as COMET \cite{COMET2018} and Mu2e \cite{Bartoszek2015}, could explore the $\mu\to eX$ process at the same level as the past experiments using free muons.

One expected advantage of muonic atoms is to evade the background problem we mentioned above.
The signal energy is monochromatic in the decay of a free muon, while the electron energy spectrum in the decay of a muon in orbit has a finite width because of the nuclear recoil.
This fact allows us to search for the signal in a preferable energy region where the signal-to-background ratio is large.
In the special case of a small $m_X$, the maximum energy of the signal is close to the signal energy of the $\mu^-\to e^-$ conversion, which is the main topic of the COMET and Mu2e experiments.
This means that the electron detector for the $\mu^-\to e^-$ conversion is also optimized for the $\mu^-\to e^-X$ search.
Thus the searches for $\mu^-\to e^-X$ using muonic atoms will be complementary to searches using free muon decays.

Another merit of muonic atoms is that the shape and the nuclear dependence of the electron spectrum are available to obtain detailed information on new physics.
The model identification by measuring such characteristic observables has been discussed in another lepton-flavor-violating process, $\mu^-e^-\to e^-e^-$ in a muonic atom \cite{Uesaka2016,Uesaka2018,Kuno2019}.
Despite its importance, no one has studied the model dependence of observables in the $\mu^-\to e^-X$ process.

Our goal of this article is to understand the model-discriminating power of the $\mu^-\to e^-X$ process in a muonic atom.
For a simple discussion of the model dependence, we introduce three effective models in Sec.~\ref{sec:Formalism}.
Then, we formulate the rate of $\mu^-\to e^-X$ in a nuclear Coulomb potential.
In Sec.~\ref{sec:Results}, we show numerical results and discuss the model dependence of observables.
Finally, we summarize this article in Sec.~\ref{sec:Summary}.

\section{Formulation \label{sec:Formalism}}

In this section, we formulate the spectrum of an emitted electron from the $\mu^-\to e^-X$ process in a muonic atom.
Here, we assume a boson $X$ lighter than muons.
To investigate the model dependence, we consider three simple effective models, called $S_0$, $S_1$, and $V_1$, which are defined as follows.

First, we assume that $X$ is a scalar field and the effective interaction Lagrangian to charged leptons is given as
\begin{align}
\mathcal{L}_{S_0}=& X\overline{e}\left(g_L^{S_0}P_L+g_R^{S_0}P_R\right)\mu+[\text{H.c.}],
\label{eq:Lagrangian_yukawa}
\end{align}
where $P_{L/R}=\left(1\mp\gamma_5\right)/2$ is a projection operator, and $g_{L/R}^{S_0}$ are dimensionless coupling constants.
In this article, we do not consider how to make the model UV complete, and we write down only the relevant part of the Lagrangian.
This type of Lagrangian was also analyzed in Refs.~\cite{Tormo2011,Heeck2017}.
In this model, keeping an electron mass $m_e=0.510999$ MeV, we find the rate of the exotic free muon decay $\mu\to eX$ to be
\begin{align}
\Gamma_0=&\frac{m_\mu}{32\pi}\sqrt{\lambda\left(1,r_e^2,r_X^2\right)}\left\{\left(\left|g_L^{S_0}\right|^2+\left|g_R^{S_0}\right|^2\right)\left(1-r_X^2+r_e^2\right)+4r_e\mathrm{Re}\left[g_L^{S_0}g_R^{S_0*}\right]\right\},
\label{eq:gamma_free_s0}
\end{align}
where $r_X=m_X/m_\mu$, $r_e=m_e/m_\mu$, and $\lambda\left(x,y,z\right)=x^2+y^2+z^2-2xy-2yz-2zx$.
Multiplying it with the lifetime of muon $\tau_\mu=192\pi^3/(G_F^2m_\mu^5)$, where $G_F=1.166\times 10^{-5}$ GeV$^{-2}$ is the Fermi coupling constant, we obtain the branching ratio for the free muon, $Br\left(\mu\to eX\right)=\tau_\mu\Gamma_0$.
For reference, suppose that $g_L^{S_0}=g_R^{S_0}$($=g^{S_0}$) and $m_X=0$.
Then, using $Br<2.6\times 10^{-6}$ \cite{Jodidio1986}, we obtain the constraint for the coupling constant,
\begin{align}
\left|g^{S_0}\right|^2<3.7\times 10^{-22}.
\end{align}

Second, we assume the following derivative coupling for the scalar $X$,
\begin{align}
\mathcal{L}_{S_1}=&(-i)\frac{\partial^\alpha X}{\Lambda_{S_1}}\overline{e}\gamma_\alpha\left(g_L^{S_1}P_L+g_R^{S_1}P_R\right)\mu+[\text{H.c.}],
\label{eq:Lagrangian_derivative}
\end{align}
where $\Lambda_{S_1}$ is an arbitrary energy scale to keep coupling constants $g_{L/R}^{S_1}$ dimensionless.
The rate of the free muon decay is given as
\begin{align}
\Gamma_0=&\frac{m_\mu}{32\pi}\sqrt{\lambda\left(1,r_e^2,r_X^2\right)}\left(\frac{m_\mu}{\Lambda_{S_1}}\right)^2\left\{\left(\left|g_L^{S_1}\right|^2+\left|g_R^{S_1}\right|^2\right)\left\{\left(1-r_e^2\right)^2-r_X^2\left(1+r_e^2\right)\right\}+4r_er_X^2\mathrm{Re}\left[g_L^{S_1}g_R^{S_1*}\right]\right\}.
\label{eq:gamma_free_s1}
\end{align}
Now we mention that, when both leptons are free and on mass shell, Eq.~(\ref{eq:Lagrangian_derivative}) is effectively equivalent to Eq.~(\ref{eq:Lagrangian_yukawa}) due to the Dirac equation, $\left(i\Slash{\partial}-m\right)\psi=0$.
Here, we have the relation of coupling constants given as
\begin{align}
g_{L/R}^{S_0}=&\frac{1}{\Lambda_{S_1}}\left(m_\mu g_{R/L}^{S_1}-m_eg_{L/R}^{S_1}\right).
\label{eq:coupling_relation}
\end{align}
Applying the relation, we easily prove the equality of Eqs.~(\ref{eq:gamma_free_s0}) and (\ref{eq:gamma_free_s1}).
However, Eq.~(\ref{eq:coupling_relation}) no longer holds in a Coulomb potential.
For the process in a muonic atom, it is worth investigating the quantitative differences of the observables between the two models.

Third, in addition to the scalar cases, we consider another case where $X$ is a vector field and the effective interaction is given as
\begin{align}
\mathcal{L}_{V_1}=\frac{X^{\alpha\beta}}{2\Lambda_{V_1}}\overline{e}\sigma_{\alpha\beta}\left(g_L^{V_1}P_L+g_R^{V_1}P_R\right)\mu+[\text{H.c.}],
\label{eq:Lagrangian_dipole}
\end{align}
where $X^{\alpha\beta}=\partial^\alpha X^\beta-\partial^\beta X^\alpha$ is the field strength of the $X$.
The couplings $g_{L/R}^{V_1}$ are dimensionless again due to the arbitrary scale $\Lambda_{V_1}$.
As with the previous models, the decay rate for the free muon is given as
\begin{align}
\Gamma_0=&\frac{m_\mu}{32\pi}\sqrt{\lambda\left(1,r_e^2,r_X^2\right)}\left(\frac{m_\mu}{\Lambda_{V_1}}\right)^2\left\{\left(\left|g_L^{V_1}\right|^2+\left|g_R^{V_1}\right|^2\right)\left\{2-r_X^2-r_X^4-r_e^2\left(4+r_X^2\right)+2r_e^4\right\}-12r_er_X^2\mathrm{Re}\left[g_L^{V_1}g_R^{V_1*}\right]\right\}.
\end{align}

Next, we formulate the rate of the $\mu^-\to e^-X$ process in a muonic atom.
We assume the independent particle model of a muonic atom and an initial muon in a $1s$ orbit.
We define the transition amplitude $\mathcal{M}$ as
\begin{align}
2\pi\delta(E_X+E_e-m_\mu^*)\mathcal{M}=\Braket{e^-_{p_e},X_{p_X}|\int d^4x\mathcal{L}_M|\mu^-_{1s}},
\end{align}
where we take only the leading order of effective interaction.
For simplicity, we omit the spin indices.
Here, $E_X$ and $E_e$ are the energies of the emitted $X$ and electron in the final state, respectively.
$m_\mu^*=m_\mu-B_{\mu N}^{1s}$ indicates the energy of the bound muon, where $B_{\mu N}^{1s}$ is the binding energy between the nucleus and muon in a $1s$ state.
The $\mathcal{M}$ connects to the decay rate by
\begin{align}
d\Gamma=\frac{d^3p_e}{(2\pi)^32E_e}\frac{d^3p_X}{(2\pi)^32E_X}(2\pi)\delta\left(E_X+E_e-m_\mu^*\right)\frac{1}{2}\sum_{\text{spins}}\left|\mathcal{M}\right|^2.
\label{eq:decayrate}
\end{align}
The factor of $1/2$ comes from the spin average of the initial bound muon.

The transition amplitude $\mathcal{M}$ includes the overlap integrals of lepton wave functions that are solutions of the Dirac equation with the nuclear Coulomb potential \cite{rose1961}.
In the central force system, one can represent the wave function of the bound muon as
\begin{align}
\psi_\mu^{s_\mu}(\bm{r})=
\begin{pmatrix}
G(r)\chi_{-1}^{s_\mu}(\hat{r}) \\
iF(r)\chi_{+1}^{s_\mu}(\hat{r})
\end{pmatrix},
\label{eq:bound_state}
\end{align}
with a normalization condition
\begin{align}
\int d^3r\overline{\psi}_\mu^{s'_\mu}(\bm{r})\psi_\mu^{s_\mu}(\bm{r})=&\delta_{s_\mu,s'_\mu}.
\end{align}
The angular parts $\chi$ are two-component spinors, which are determined analytically.
Furthermore, we obtain the radial part and the binding energy by solving an eigenvalue problem for the radial Dirac equations,
\begin{align}
\frac{d}{dr}
\begin{pmatrix}
G(r) \\
F(r)
\end{pmatrix}
=&
\begin{pmatrix}
0 & E_\mu+m_{\mu N}+eV_{\rm{C}}(r) \\
-E_\mu+m_{\mu N}-eV_{\rm{C}}(r) & -2/r
\end{pmatrix}
\begin{pmatrix}
G(r) \\
F(r)
\end{pmatrix}.
\label{eq:Dirac_eq_for_bm}
\end{align}
The nuclear Coulomb potential $V_C$ in the equations is given as
\begin{align}
V_C(r)=\int_{0}^{\infty}dr'r'^{2}\left[\theta\left(r-r'\right)\frac{1}{r}+\theta\left(r'-r\right)\frac{1}{r'}\right]\rho\left(r'\right),
\end{align}
with a nuclear charge density $\rho(r)$.
Here, we use the reduced mass $m_{\mu N}=m_\mu m_N/(m_N+m_\mu)$ with a nuclear mass $m_N$.
After obtaining the solution where $E_\mu$ is minimized, we determine the binding energy of the $1s$ state by $B_{\mu N}^{1s}=m_{\mu N}-E_\mu$ \cite{Grotch1969}.

For the electron in the final state, it is convenient to use the multipole expansion of the state with momentum $\bm{p}_e$.
The electron scattering state with the incoming boundary condition is expressed as follows:
\begin{align}
\psi_{e,\bm{p}_e}^{s_e}\left(\bm{r}\right)=&\sum_{\kappa,\nu,m}4\pi i^{l_\kappa}(l_\kappa,m,1/2,s_e|j_\kappa,\nu)Y_{l_\kappa}^{m*}(\hat{p}_e)e^{-i\delta_\kappa}
\begin{pmatrix}
g_{E_e}^\kappa(r)\chi_\kappa^\nu(\hat{r}) \\
if_{E_e}^\kappa(r)\chi_{-\kappa}^\nu(\hat{r})
\end{pmatrix},
\label{eq:scattering_state}
\end{align}
with the Clebsch-Gordan coefficients $(l_\kappa,m,1/2,s_e|j_\kappa,\nu)$ and spherical harmonics $Y_{l_\kappa}^{m}(\hat{p}_e)$.
Here, $\kappa$ is a nonzero integer to label partial waves.
For the index $\kappa$, the total angular momentum $j_\kappa$ and the orbital angular momentum $l_\kappa$ are determined by
\begin{align}
j_\kappa=&\left|\kappa\right|-\frac{1}{2}, \\
l_\kappa=& j_\kappa+\frac{1}{2}\frac{\kappa}{|\kappa|}.
\end{align}
$\delta_\kappa$ is the phase shift of a partial wave labeled by $\kappa$.
To obtain the radial wave functions for a given $E_e$ and $\kappa$, we solve
\begin{align}
\frac{d}{dr}
\begin{pmatrix}
g^\kappa_{E_e}(r) \\
f^\kappa_{E_e}(r)
\end{pmatrix}
=&
\begin{pmatrix}
-\left(1+\kappa\right)/r & E_e+m_{eN}+eV_{\rm{C}}(r) \\
-E_e+m_{eN}-eV_{\rm{C}}(r) & -\left(1-\kappa\right)/r
\end{pmatrix}
\begin{pmatrix}
g^\kappa_{E_e}(r) \\
f^\kappa_{E_e}(r)
\end{pmatrix}.
\label{eq:Dirac_eq_for_se}
\end{align}
The normalization is taken to be
\begin{align}
\int d^3r\overline{\psi}_{e,\bm{p}'_e}^{s'_e}\left(\bm{r}\right)\psi_{e,\bm{p}_e}^{s_e}\left(\bm{r}\right)=& 2E_e\left(2\pi\right)^3\delta^{(3)}\left(\bm{p}'_e-\bm{p}_e\right)\delta_{s,s'}.
\end{align}

Using the expressions of the effective interactions, we find that the electron spectra for the three models ($M=S_0,S_1,V_1$) are universally represented as
\begin{align}
\frac{d\Gamma}{dE_e}=&\frac{\sqrt{E_e^2-m_e^2}\sqrt{E_X^2-m_X^2}}{16\pi^2}\sum_{\kappa}\left(2j_{\kappa}+1\right)\left\{\left(\left|g_L^{M}\right|^2+\left|g_R^{M}\right|^2\right)\left(P_\kappa^{M}+\overline{P}_\kappa^{M}\right)+2\mathrm{Re}\left[g_L^{M*}g_R^{M}\right]\left(P^{M}_\kappa-\overline{P}^{M}_\kappa\right)\right\},
\label{eq:spectrum}
\end{align}
where $E_X$ is a function of $E_e$ determined by the energy conservation.
To take into account nuclear recoil through $E_X$, we apply the well-known prescription as follows \cite{Tormo2011,Shanker1982,Czarnecki2011}:
\begin{align}
E_X=m_\mu^*-E_e\to m_\mu^*-E_e-\frac{E_e^2}{2m_N}.
\end{align}
This additional term represents the kinetic energy of the recoiled nucleus, and the term is sizable only at high $E_e$ but negligible at low $E_e$.
Thus, even though we do not completely consider the nuclear motion, we believe that this treatment yields a good approximation for any $E_e$.

After straightforward calculation, we obtain the explicit formulas for $P^{M}_\kappa$ and $\overline{P}^{M}_\kappa$.
For $M=S_0$, it is found that
\begin{align}
P^{S_0}_\kappa=&\left|I_{gG}^{\kappa,(l_{\kappa})}-I_{fF}^{\kappa,(l_{\kappa})}\right|^2, \\
\overline{P}^{S_0}_\kappa=&\left|I_{fG}^{\kappa,(l_{-\kappa})}+I_{gF}^{\kappa,(l_{-\kappa})}\right|^2.
\end{align}
Here, we define the overlap integral, $I_{hH}^{\kappa,(L)}$ ($h=g,f$ and $H=G,F$), as
\begin{align}
I_{hH}^{\kappa,(L)}=\int_{0}^{\infty}drr^2j_{L}\left(\sqrt{E_X^2-m_X^2}r\right)h_{E_e}^\kappa(r)H(r),
\end{align}
where $j_{l}$ is the $l$th-order spherical Bessel function.
$h$ indicates the radial wave function of the scattering electron, and $H$ indicates that of the bound muon.
This formula for $S_0$ is consistent with that in Ref.~\cite{Tormo2011}.
More complicated expressions for $P_\kappa^{S_1}$, $\overline{P}_\kappa^{S_1}$, $P_\kappa^{V_1}$, and $\overline{P}_\kappa^{V_1}$ are given in the Appendix.

If we neglect the electron mass, we find that the components of the transition probability satisfy
\begin{align}
\overline{P}_{-\kappa}^M=P_{\kappa}^M,
\end{align}
which is valid regardless of $M$.
Because of this symmetry, the cross term between $g_L^M$ and $g_R^M$ disappears after summing over $\kappa$.
This observation is understandable because the interference between left- and right-handed components should vanish for the final electron if $m_e=0$.

The end point energy $E_\mathrm{end}^{m_X}$ of the electron spectrum is kinematically determined as
\begin{align}
E_\mathrm{end}^{m_X}=&\frac{\left(m_N+m_\mu^*-m_X\right)^2-m_N^2+m_e^2}{2\left(m_N+m_\mu^*-m_X\right)},
\label{eq:endpoint_energy}
\end{align}
which is obtained by solving the relativistic relation of the energy-momentum conservation.
Approximately, Eq.~(\ref{eq:endpoint_energy}) is represented by
\begin{align}
E_\mathrm{end}^{m_X}\simeq& m_\mu^*-m_X-\frac{\left(m_\mu^*-m_X\right)^2}{2m_N},
\end{align}
where the third term is interpreted as the kinetic energy of the recoiled nucleus.

\section{Numerical Results \label{sec:Results}}

To obtain the radial wave functions of charged leptons and the binding energy of a muonic atom, we solve the differential equations Eq.~(\ref{eq:Dirac_eq_for_bm}) for the initial muon and Eq.~(\ref{eq:Dirac_eq_for_se}) for the final electron.
In solving the differential equations, we use the fourth-order Runge-Kutta method.
The correctness of our calculation code is numerically checked by comparing it with the analytic result for a point-charge density.

For reference, we focus on two kinds of nuclei as a target material.
One is aluminum $^{27}$Al, which will be used in the coming COMET and Mu2e experiments.
The other is gold $^{197}$Au, which was used in the SINDRUM II experiment \cite{Bertl2006}.
For both nuclei, we assume the two-parameter Fermi distribution as the nuclear charge density, given as
\begin{align}
\rho(r)=\frac{Ze}{4\pi}\frac{\rho_0}{1+\exp\left(\dfrac{r-r_0}{a}\right)},
\label{eq:rho}
\end{align}
where $Z$ is the proton number of the target nucleus, and $e$ is the magnitude of the elementary charge.
The parameters of the distribution $r_0$ and $a$ are given in Table \ref{tab:parameters}, and $\rho_0$ is a normalization factor.
By solving Eq.~(\ref{eq:Dirac_eq_for_bm}), we obtain the values of the binding energy $B_{\mu N}^{1s}$ shown in Table \ref{tab:parameters}.
Substituting the binding energy into Eq.~(\ref{eq:endpoint_energy}), we find the end point energy $E_\mathrm{end}^{m_X}$ for an arbitrary $m_X$.
The values of $E_\mathrm{end}^{m_X}$ for $m_X=0$, $25$, $50$ MeV are shown in Table~\ref{tab:energies}.
\begin{table}[htb]
\setlength{\extrarowheight}{3pt}
\centering
 \caption{Parameters for each nucleus and the calculated energies. The forth and fifth columns are the parameters in Eq.~(\ref{eq:rho}), given by Ref.~\cite{Jager1974}. The sixth and seventh columns are $E_\mu$ and $B_{\mu N}^{1s}$ obtained by our calculation.}
 \vspace{0.4cm}
 \doublerulesep 0.8pt \tabcolsep 0.4cm
  \begin{tabular}{cccccccc} \hline\hline
    Nuclei & $Z$ & $A$ & $m_N$ (MeV) & $r_0$ (fm) & $a$ (fm) & $E_\mu$ (MeV) & $B_{\mu N}^{1s}$ (MeV) \\ \hline
    $^{27}$Al & 13 & 27 & 25133 & 2.845 & 0.569 & 104.75 & 0.4629 \\
    $^{197}$Au & 79 & 197 & 183473 & 6.38 & 0.535 & 95.48 & 10.12 \\ \hline\hline
  \end{tabular}
  \label{tab:parameters}
\end{table}
\begin{table}[htb]
\setlength{\extrarowheight}{3pt}
\centering
 \caption{End point energies $E_\mathrm{end}^{m_X}$ for $m_X=0$, $25$, $50$ MeV.}
 \vspace{0.4cm}
 \doublerulesep 0.8pt \tabcolsep 0.4cm
  \begin{tabular}{cccc} \hline\hline
    Nuclei & $E_\mathrm{end}^{0}$ (MeV) & $E_\mathrm{end}^{25\mathrm{MeV}}$ (MeV) & $E_\mathrm{end}^{50\mathrm{MeV}}$ (MeV) \\ \hline
    $^{27}$Al & 104.98 & 80.07 & 55.13 \\
    $^{197}$Au & 95.51 & 70.52 & 45.53 \\ \hline\hline
  \end{tabular}
  \label{tab:energies}
\end{table}

\begin{figure}[htb]
  \centering
    \begin{tabular}{c}
      % 1
      \begin{minipage}{0.45\hsize}
		\centering
          \includegraphics[clip, width=7.0cm]{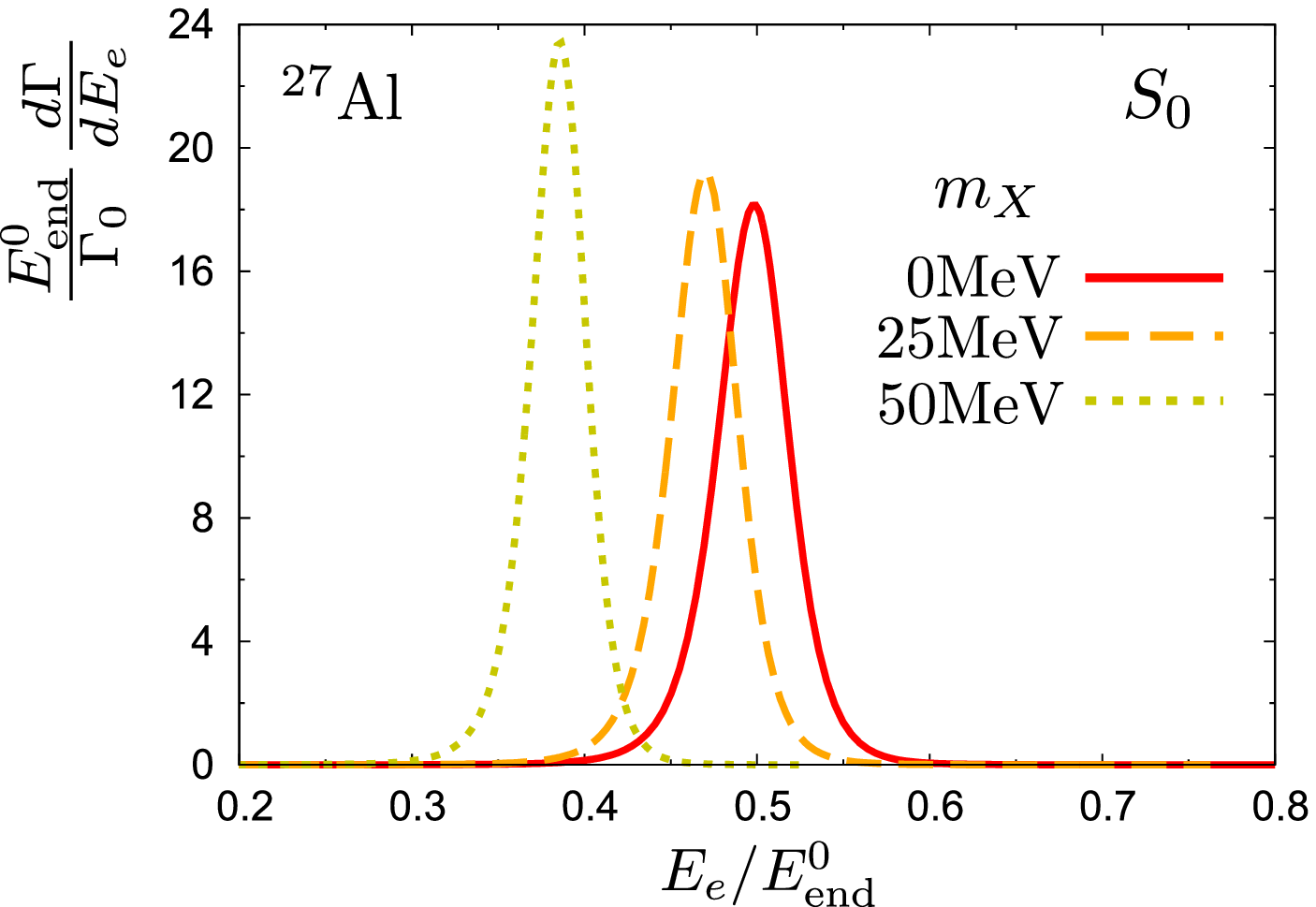}
          \\ (a)
      \end{minipage}%
      % 2
      \begin{minipage}{0.45\hsize}
        \centering
          \includegraphics[clip, width=7.0cm]{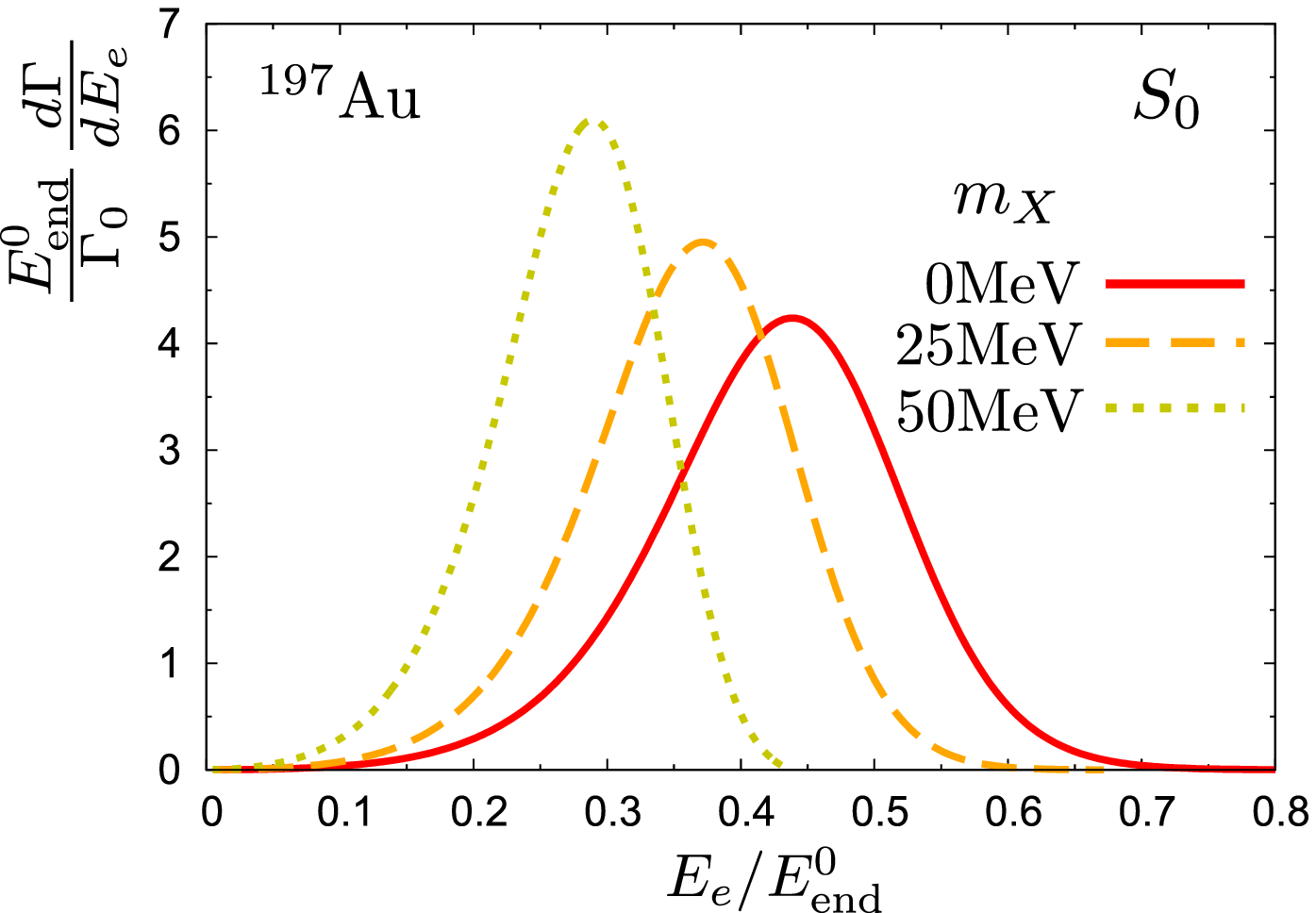}
          \\ (b)
      \end{minipage}%
    \end{tabular}
    \caption{
    Spectra of the emitted electron.
    (a) is for $^{27}$Al and (b) is for $^{197}$Au.
    The horizontal axis is the electron energy $E_e$ divided by its maximum energy $E_\mathrm{end}^0$.
    }
    \label{fig:spectrum_Al_Au}
\end{figure}

Figure~\ref{fig:spectrum_Al_Au}(a) shows the electron spectra for the aluminum nucleus.
The spectra are normalized by the rate for a free muon, whose expression for each model is given in Sec.~\ref{sec:Formalism}.
Here, we plot only the spectrum of $S_0$ model because the differences between the models are too small to recognize in this energy scale.
Each curve in Fig.~\ref{fig:spectrum_Al_Au} corresponds to $m_X$, where the electron energy is universally normalized by the end point energy for massless $X$, $E_\mathrm{end}^0$.
As well as the end point energy, the position of the spectrum peak is shifted lower as $m_X$ increases.
The peak position is approximately given as $E_e\simeq(m_\mu^2-m_X^2)/(2m_\mu)$, which is the expected signal energy if the momentum of the initial muon is assumed to be zero.
We also note that the spectrum for $^{197}$Au shown in Fig.~\ref{fig:spectrum_Al_Au}(b) has a larger width than $^{27}$Al.
This is because the momentum uncertainty of the initial muon is larger as the nucleus has a stronger Coulomb field.

Here, we suppose that the search for the $\mu^-\to e^-X$ process would be performed in an experiment for the $\mu^-\to e^-$ conversion, which is optimized to detect high-energy electrons.
In this case, the search would be sensitive to the high-energy shape of the spectrum.
Hereafter, we set $m_X=0$ as a reference because the high-energy end point of $\mu^-\to e^-X$ is close to the signal energy of the $\mu^-\to e^-$ conversion.

\begin{figure}[htb]
  \centering
    \begin{tabular}{c}
      % 1
      \begin{minipage}{0.45\hsize}
		\centering
          \includegraphics[clip, width=7.0cm]{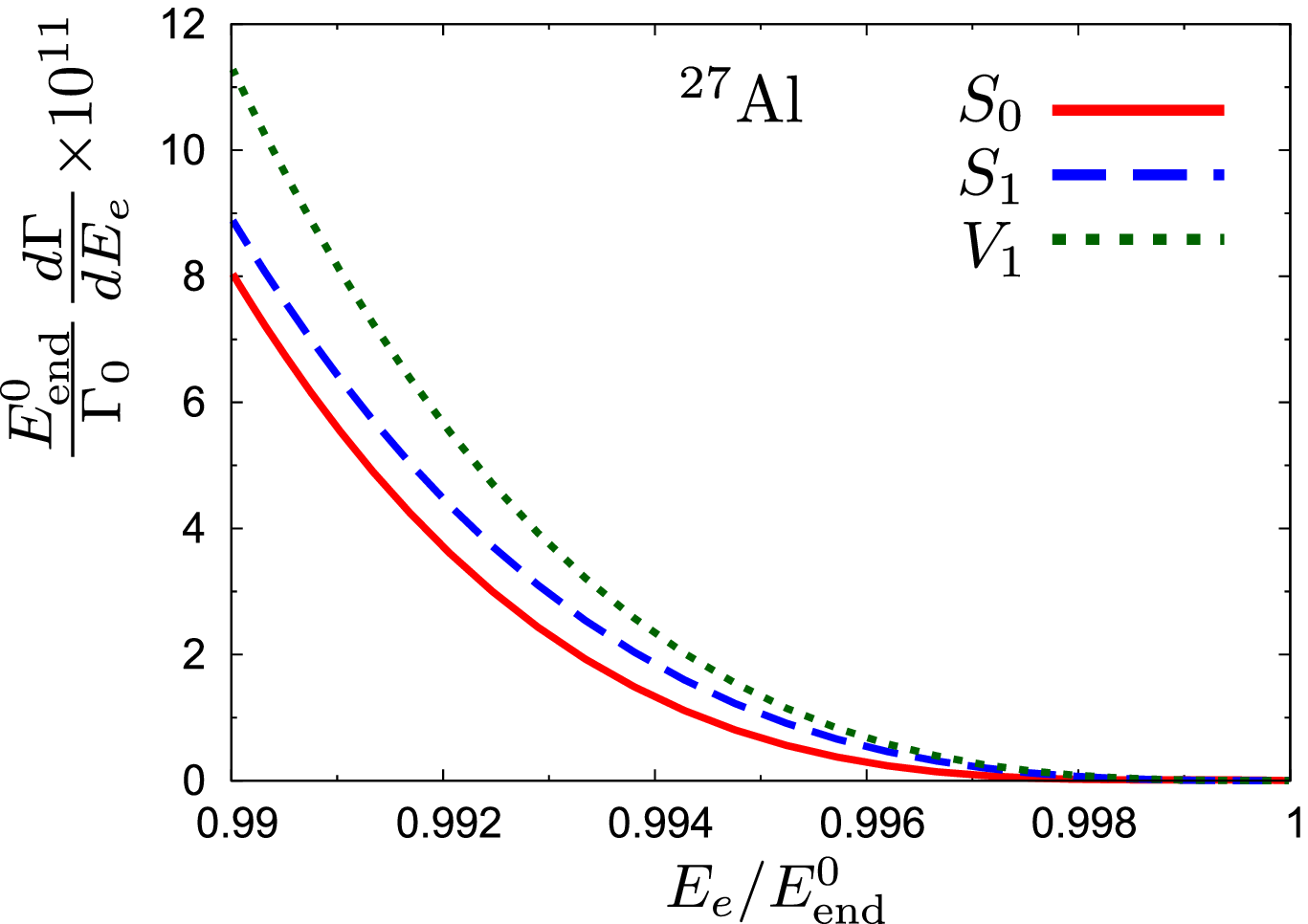}
          \\ (a)
      \end{minipage}%
      % 2
      \begin{minipage}{0.45\hsize}
        \centering
          \includegraphics[clip, width=7.0cm]{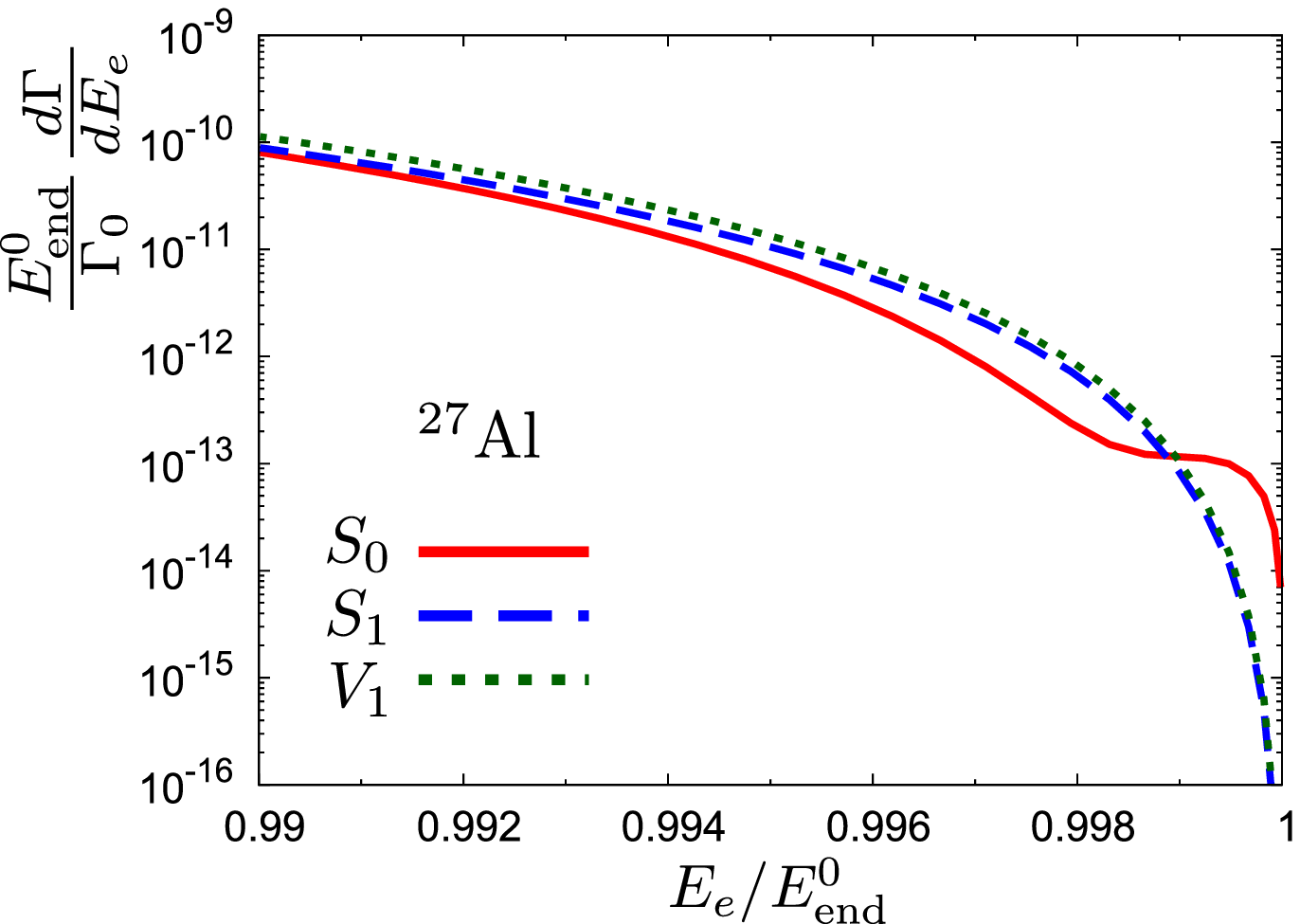}
          \\ (b)
      \end{minipage}%
    \end{tabular}
    \caption{
    Spectra of the emitted electron for $^{27}$Al.
	The region that $0.99\le E_e/E^0_\mathrm{end}\le 1$ is shown.
    On the y axis, we use a linear scale in (a) and a logarithmic scale in (b).
    }
    \label{fig:spectrum_Al}
\end{figure}

We plot the spectra for $^{27}$Al in the range of $0.99\le E_e/E_\mathrm{end}^{0}\le 1$ in Fig.~\ref{fig:spectrum_Al}.
In this figure, one can recognize the difference between the models of the boson $X$.
In particular, the high-energy tail of the $V_1$ model indicated by the dotted (green) curve is larger than the others.
This observation suggests that the analysis of the end point spectrum is more sensitive to the $V_1$ model than the others.

We should comment on the spectrum for the $S_0$ model shown by the solid (red) curve in Fig.~\ref{fig:spectrum_Al}.
One may find that the spectrum for the $S_0$ model is unnaturally suppressed near $E_e/E_\mathrm{end}^0\simeq 0.998$, which is clearly seen in Fig.~\ref{fig:spectrum_Al}(b).
This happens due to the following three facts:
First, the spectrum is dominated by the contribution of $\kappa=-1$ in Eq.~(\ref{eq:spectrum}).
Second, $P_{-1}^{S_0}$ vanishes when $E_e=E_\mu$ \cite{Shanker1982}.
Third, $E_\mu$ is slightly smaller than $E_\mathrm{end}^0$ due to the finite nuclear mass.
Organizing them, we notice that the main contribution of the spectrum vanishes at $E_e=E_\mu\lesssim E_\mathrm{end}^0$, which is close to but smaller than $E_\mathrm{end}^0$.
This interesting property characterizes the $S_0$ model.
In practice, after the confirmation of $X$, we need much more careful measurement to identify the spectrum shape near the end point.

This characteristic feature of the $S_0$ model was not reported in the previous study~\cite{Tormo2011}, where the original muon mass $m_\mu$ seemed to be used instead of the reduced mass $m_{\mu N}$ in the calculation of the binding energy $B_{\mu N}$.
If one calculates $B_{\mu N}$ ignoring the nuclear mass, one finds that $E_\mu$ is larger than $E_\mathrm{end}^0$, and therefore, the disappearance of $P_{-1}^{S_0}$ discussed above does not happen for any physical $E_e$.
Here, we emphasize that the characteristic feature can be only seen with the nuclear mass taken to be finite.

\begin{figure}[htb]
  \centering
    \begin{tabular}{c}
      % 1
      \begin{minipage}{0.45\hsize}
		\centering
          \includegraphics[clip, width=7.0cm]{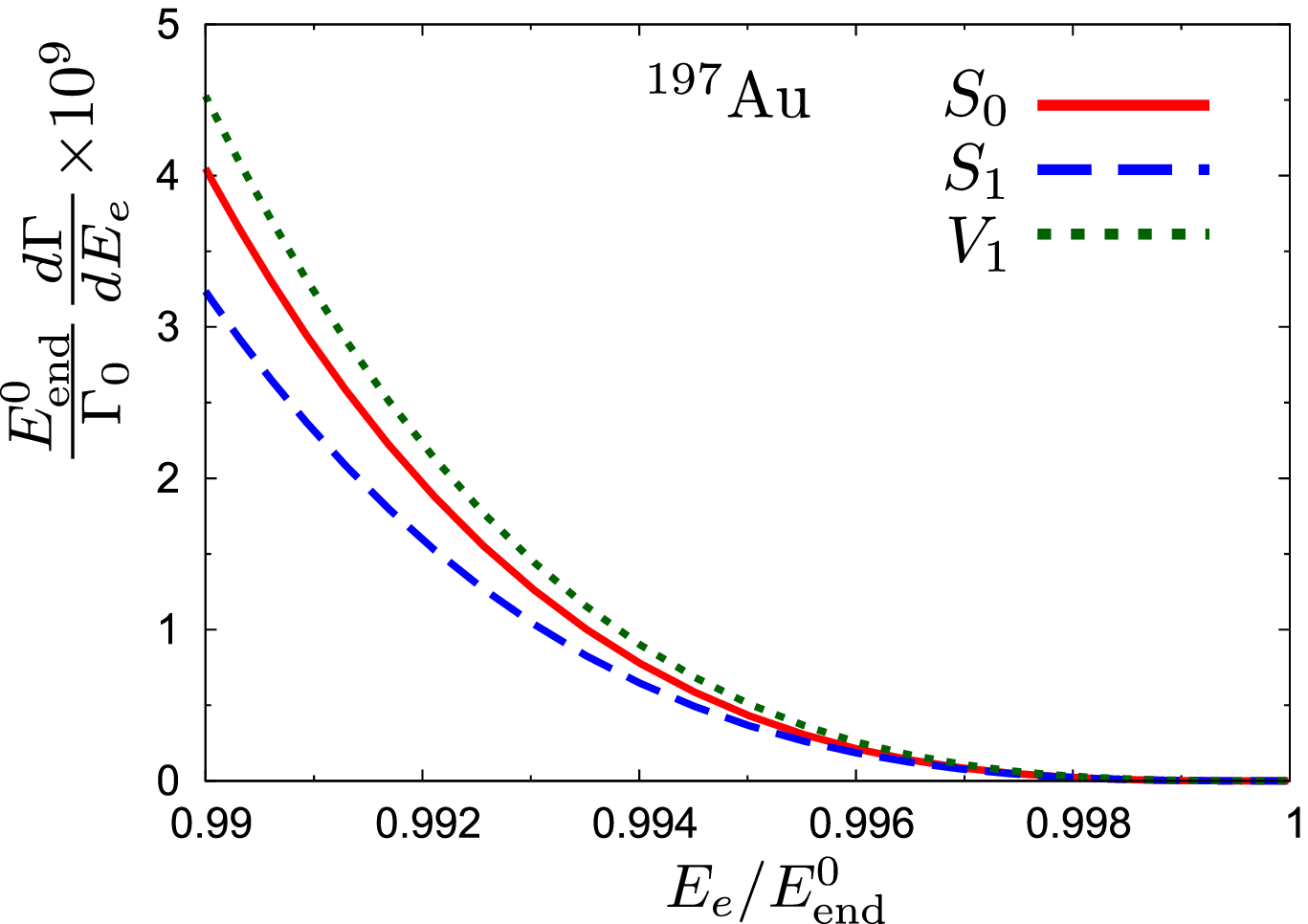}
          \\ (a)
      \end{minipage}%
      % 2
      \begin{minipage}{0.45\hsize}
        \centering
          \includegraphics[clip, width=7.0cm]{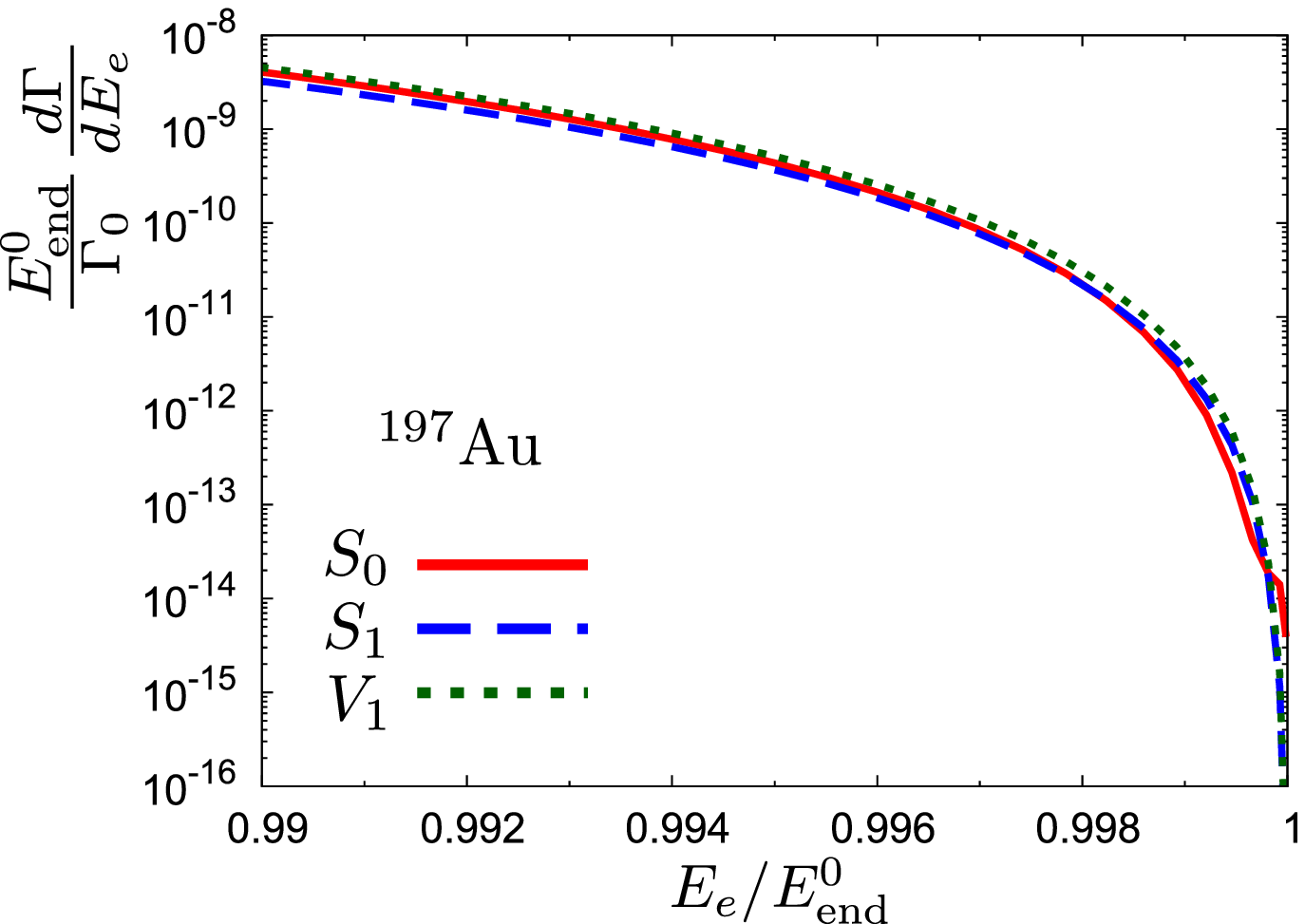}
          \\ (b)
      \end{minipage}%
    \end{tabular}
    \caption{
    Spectra of the emitted electron for $^{197}$Au.
    The region that $0.99\le E_e/E^0_\mathrm{end}\le 1$ is shown.
    On the y axis, we use a linear scale in (a) and a logarithmic scale in (b).
    }
    \label{fig:spectrum_Au}
\end{figure}

Also, Fig.~\ref{fig:spectrum_Au} shows the spectrum for $^{197}$Au in the range of $0.99\le E_e/E_\mathrm{end}^{0}\le 1$.
We find that the high-energy tail is much larger than $^{27}$Al.
As with $^{27}$Al, the tail of the $V_1$ model is the largest of the three models.
We cannot recognize the suppression of the spectrum near the end point for the $S_0$ model in the $^{27}$Al case, because the nuclear mass $m_N$ is so heavy that $E_\mu$ is sufficiently close to $E_\mathrm{end}^0$.

\begin{figure}[htb]
	\centering
    \includegraphics[clip, width=7.0cm]{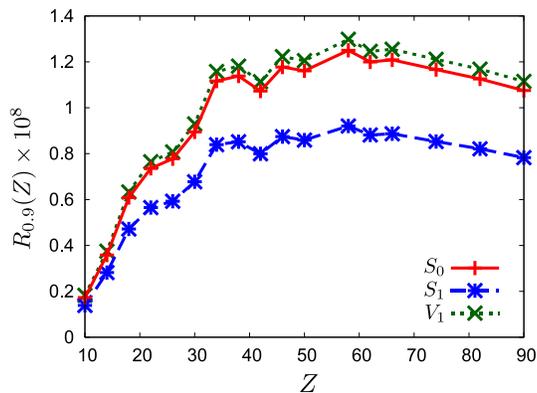}
    \caption{
    The $Z$ dependence of $R_{0.9}(Z)$ defined in Eq.~(\ref{eq:Rz}).
    Sampled points are shown by crosses.
    For the simplicity of the calculation, we use the uniform distribution with the nuclear radius of $1.2A^{1/3}$fm as the nuclear charge density.
    We take the mass number $A$ of the most abundant isotope for each $Z$ \cite{Berglund2011}.
    }
    \label{fig:m2eX_e0.9}
\end{figure}

Finally, we discuss which nucleus is preferable for the $\mu^-\to e^-X$ search.
Suppose that the new physics search using muonic atoms is performed by measuring the number of electrons with an energy close to the signal energy of $\mu^-\to e^-$ conversion, which is equal to $E_\mathrm{end}^0$.
We define a net branching ratio as
\begin{align}
Br_x(Z)=&\tilde{\tau}_\mu\int_{x}^{1}d\left(\frac{E_e}{E_\mathrm{end}^{0}}\right)E_\mathrm{end}^{0}\frac{d\Gamma}{dE_e},
\end{align}
where $\tilde{\tau}$ is the lifetime of a muonic atom listed in Ref.~\cite{Suzuki1987}.
This value corresponds to the number of electrons with $E_e\ge xE_\mathrm{end}^0$ ($x<1$) coming from $\mu^-\to e^-X$ normalized by the created number of muonic atoms.
For further convenience, we define
\begin{align}
R_x(Z)=\frac{\tilde{\tau}_\mu}{\tau_\mu}\int_{x}^{1}d\left(\frac{E_e}{E_\mathrm{end}^{0}}\right)\frac{E_\mathrm{end}^{0}}{\Gamma_0}\frac{d\Gamma}{dE_e},
\label{eq:Rz}
\end{align}
so that
\begin{align}
Br_x(Z)=& R_x(Z)Br\left(\mu^+\to e^+X\right).
\label{eq:Br_x=RBr}
\end{align}
Setting $x=0.9$, we find that the $Z$ dependence of $R_{0.9}(Z)$ is shown in Fig.~\ref{fig:m2eX_e0.9}.
One can see that the typical value of $R_{0.9}(Z)$ is $O(10^{-9}-10^{-8})$.
As larger nuclei, the lifetime of muonic atoms is shorter, but the high-energy tail of the electron spectrum gets larger.
Because of the cancellation of the two effects \cite{Tormo2011}, the $Z$ dependence of $R$ is not so strong above $Z\approx 30$.
Considering the current experimental constraint of $Br\left(\mu^+\to e^+X\right)$, we find that the current upper limit of the net branching ratio is $Br_{0.9}(Z)<O(10^{-15}-10^{-14})$.
Since the goal of the created number of muons in the planned $\mu^-\to e^-$ conversion searches \cite{COMET2018,Bartoszek2015} is $\mathcal{O}\left(10^{18}\right)$, it would be possible to reach the constraint by the near-future muon sources.

\section{Summary \label{sec:Summary}}

We have investigated the $\mu^-\to e^-X$ process in muonic atoms as an interesting candidate to constrain the property of light neutral bosons.
Assuming three simple effective models of the unknown boson, we have discussed the model dependence of the electron spectrum.
As a result, we have found that the spectrum near the end point strongly depends on the property of the boson $X$.
We have also shown that the nuclear dependence of the net branching ratio is moderate.

A remaining theoretical problem is to include radiative corrections in the calculation for the spectrum near the high-energy end point, which is shown to be important for ordinary decay of a muon in orbit \cite{Szafron2016}.
Although we need further studies of the realistic sensitivity of experiments, we believe that careful measurements for the electron spectrum in a muon decay are useful to find unknown invisible bosons and to identify their property.

\begin{acknowledgments}
We thank Y.~Kuno, C.~Wu, T.~Xing, J.~Sato, and T.~Sato for fruitful comments.
This work was supported by JSPS KAKENHI Grant Number JP18H01210 and the Sasakawa Scientific Research Grant from the Japan Science Society.
\end{acknowledgments}

\appendix

\section{Full expressions of the transition provability \label{app:full_expression}}

We show the expressions for $P_\kappa^M$ and $\overline{P}_\kappa^M$ ($M=S_1,V_1$).
For the $S_1$ model,
\begin{align}
P_\kappa^{S_1}=&\frac{E_X^2}{\Lambda_{S_1}^2}\left|I_{gG}^{\kappa,(l_\kappa)}+I_{fF}^{\kappa,(l_\kappa)}+\frac{\sqrt{E_X^2-m_X^2}}{E_X}\left(\frac{2+l_\kappa+\kappa}{2l_\kappa+1}I_{gF}^{\kappa,(l_\kappa+1)}-\frac{l_\kappa-\kappa}{2l_\kappa+1}I_{fG}^{\kappa,(l_\kappa+1)}\right.\right. \nonumber\\
&\left.\left.-\frac{l_\kappa-1-\kappa}{2l_\kappa+1}I_{gF}^{\kappa,(l_\kappa-1)}+\frac{l_\kappa+1+\kappa}{2l_\kappa+1}I_{fG}^{\kappa,(l_\kappa-1)}\right)\right|^2, \\
\overline{P}_\kappa^{S_1}=&\frac{E_X^2}{\Lambda_{S_1}^2}\left|I_{fG}^{\kappa,(l_{-\kappa})}-I_{gF}^{\kappa,(l_{-\kappa})}+\frac{\sqrt{E_X^2-m_X^2}}{E_X}\left(\frac{2+l_{-\kappa}-\kappa}{2l_{-\kappa}+1}I_{fF}^{\kappa,(l_{-\kappa}+1)}+\frac{l_{-\kappa}+\kappa}{2l_{-\kappa}+1}I_{gG}^{\kappa,(l_{-\kappa}+1)}\right.\right. \nonumber\\
&\left.\left.-\frac{l_\kappa-1+\kappa}{2l_{-\kappa}+1}I_{fF}^{\kappa,(l_{-\kappa}-1)}-\frac{l_{-\kappa}+1-\kappa}{2l_{-\kappa}+1}I_{gG}^{\kappa,(l_{-\kappa}-1)}\right)\right|^2.
\end{align}
For the $V_1$ model,
\begin{align}
P_\kappa^{V_1}=&\frac{E_X^2}{\Lambda_{V_1}^2}\left|\sqrt{\frac{l_\kappa+1}{l_\kappa}}\left(\frac{l_\kappa-1-\kappa}{2l_\kappa+1}I_{gF}^{\kappa,(l_\kappa-1)}+\frac{l_\kappa+1+\kappa}{2l_\kappa+1}I_{fG}^{\kappa,(l_\kappa-1)}\right)\right. \nonumber\\
&+\sqrt{\frac{l_\kappa}{l_\kappa+1}}\left(\frac{l_\kappa+2+\kappa}{2l_\kappa+1}I_{gF}^{\kappa,(l_\kappa+1)}+\frac{l_\kappa-\kappa}{2l_\kappa+1}I_{fG}^{\kappa,(l_\kappa+1)}\right) \nonumber\\
&\left.-\frac{\sqrt{E_X^2-m_X^2}}{E_X}\frac{1+\kappa}{\sqrt{l_\kappa\left(l_\kappa+1\right)}}\left(I_{gG}^{\kappa,(l_\kappa)}+I_{fF}^{\kappa,(l_\kappa)}\right)\right|^2 \nonumber\\
&+\frac{E_X^2}{\Lambda_{V_1}^2}\left|\frac{1-\kappa}{\sqrt{l_{-\kappa}\left(l_{-\kappa}+1\right)}}\left(I_{gF}^{\kappa,(l_{-\kappa})}-I_{fG}^{\kappa,(l_{-\kappa})}\right)\right. \nonumber\\
&+\frac{\sqrt{E_X^2-m_X^2}}{E_X}\sqrt{\frac{l_{-\kappa}+1}{l_{-\kappa}}}\left(\frac{l_{-\kappa}+1-\kappa}{2l_{-\kappa}+1}I_{gG}^{\kappa,(l_{-\kappa}-1)}-\frac{l_{-\kappa}-1+\kappa}{2l_{-\kappa}+1}I_{fF}^{\kappa,(l_{-\kappa}-1)}\right) \nonumber\\
&\left.+\frac{\sqrt{E_X^2-m_X^2}}{E_X}\sqrt{\frac{l_{-\kappa}}{l_{-\kappa}+1}}\left(\frac{l_{-\kappa}+\kappa}{2l_{-\kappa}+1}I_{gG}^{\kappa,(l_{-\kappa}+1)}-\frac{l_{-\kappa}+2-\kappa}{2l_{-\kappa}+1}I_{fF}^{\kappa,(l_{-\kappa}+1)}\right)\right|^2 \nonumber\\
&+\frac{m_X^2}{\Lambda_{V_1}^2}\left|\frac{l_\kappa-1-\kappa}{2l_\kappa+1}I_{gF}^{\kappa,(l_{\kappa}-1)}+\frac{l_\kappa+1+\kappa}{2l_\kappa+1}I_{fG}^{\kappa,(l_{\kappa}-1)}-\frac{l_\kappa+2+\kappa}{2l_\kappa+1}I_{gF}^{\kappa,(l_{\kappa}+1)}-\frac{l_\kappa-\kappa}{2l_\kappa+1}I_{fG}^{\kappa,(l_{\kappa}+1)}\right|^2,
\end{align}
\begin{align}
\overline{P}_\kappa^{V_1}=&\frac{E_X^2}{\Lambda_{V_1}^2}\left|\sqrt{\frac{l_{-\kappa}+1}{l_{-\kappa}}}\left(\frac{l_{-\kappa}+1-\kappa}{2l_{-\kappa}+1}I_{gG}^{\kappa,(l_{-\kappa}-1)}-\frac{l_{-\kappa}-1+\kappa}{2l_{-\kappa}+1}I_{fF}^{\kappa,(l_{-\kappa}-1)}\right)\right. \nonumber\\
&+\sqrt{\frac{l_{-\kappa}}{l_{-\kappa}+1}}\left(\frac{l_{-\kappa}+\kappa}{2l_{-\kappa}+1}I_{gG}^{\kappa,(l_{-\kappa}+1)}-\frac{l_{-\kappa}+2-\kappa}{2l_{-\kappa}+1}I_{fF}^{\kappa,(l_{-\kappa}+1)}\right) \nonumber\\
&\left.-\frac{\sqrt{E_X^2-m_X^2}}{E_X}\frac{1-\kappa}{\sqrt{l_{-\kappa}\left(l_{-\kappa}+1\right)}}\left(I_{gF}^{\kappa,(l_{-\kappa})}-I_{fG}^{\kappa,(l_{-\kappa})}\right)\right|^2 \nonumber\\
&+\frac{E_X^2}{\Lambda_{V_1}^2}\left|\frac{1+\kappa}{\sqrt{l_{\kappa}\left(l_{\kappa}+1\right)}}\left(I_{gG}^{\kappa,(l_{\kappa})}+I_{fF}^{\kappa,(l_{\kappa})}\right)\right. \nonumber\\
&+\frac{\sqrt{E_X^2-m_X^2}}{E_X}\sqrt{\frac{l_{\kappa}+1}{l_{\kappa}}}\left(\frac{l_{\kappa}-1-\kappa}{2l_{\kappa}+1}I_{gF}^{\kappa,(l_{\kappa}-1)}+\frac{l_{\kappa}+1+\kappa}{2l_{\kappa}+1}I_{fG}^{\kappa,(l_{\kappa}-1)}\right) \nonumber\\
&\left.+\frac{\sqrt{E_X^2-m_X^2}}{E_X}\sqrt{\frac{l_{\kappa}}{l_{\kappa}+1}}\left(\frac{l_{\kappa}+2+\kappa}{2l_{\kappa}+1}I_{gF}^{\kappa,(l_{\kappa}+1)}-\frac{l_{\kappa}-\kappa}{2l_{\kappa}+1}I_{fG}^{\kappa,(l_{\kappa}+1)}\right)\right|^2 \nonumber\\
&+\frac{m_X^2}{\Lambda_{V_1}^2}\left|\frac{l_{-\kappa}+1-\kappa}{2l_{-\kappa}+1}I_{gG}^{\kappa,(l_{-\kappa}-1)}-\frac{l_{-\kappa}-1+\kappa}{2l_{-\kappa}+1}I_{fF}^{\kappa,(l_{-\kappa}-1)}-\frac{l_{-\kappa}+\kappa}{2l_{-\kappa}+1}I_{gG}^{\kappa,(l_{-\kappa}+1)}+\frac{l_{-\kappa}+2-\kappa}{2l_{-\kappa}+1}I_{fF}^{\kappa,(l_{-\kappa}+1)}\right|^2.
\end{align}

\end{document}